# A New Method for the Discovery of the Distant Exoplanets II: Question of Signal-to-Noise Ratio

P. B. Lerner[1],

Abstract

Exoplanets with a long orbital period are difficult to discover by extant methods. Our first paper (Lerner 2023) proposed a Hanbury-Brown-Twiss (HBT) interferometry-inspired method to find exoplanets with a very slow-changing influence on their host star. Traditional HBT interferometry measured the modulus of the correlation function. Slight modification of the HBT can determine its real part. Simultaneous observation of both characteristics, e.g. in a binocular setting of telescopes, is exceptionally sensitive to the asymmetry of the luminous object. However, the issue of a very small signal-to-noise ratio (SNR) for HBT makes application of this method difficult. In the current paper, we discuss the possibilities to enhance the low SNR of the method.

[1] Retired, pblerner18@gmail.com, peter.lerner@faculty.umgc.edu.

1. **Introduction**

This paper studies the proposal to implement a binocular scheme for the combined Max-Zehnder and heterodyne interferometry. The quadratures of the intensity signal greatly elucidate the asymmetry of the correlation function of the observed compound object (star + planet). Because of the discontinuity of the phase variation of the correlation function, this method has no fundamental bounds, being restricted only by instrumental limitations.

In our first paper (Lerner 2023), we proposed a Hanbury-Brown-Twiss (HBT) interferometry-inspired method to find exoplanets with a very slow-changing influence on their host star. We chose it because it is relatively insensitive to the absolute distance mismatch between binoculars (de Graauw 1982). Our motivation was as follows.

Current methods of the discovery of exoplanets other than direct imaging are fourfold. Firstly, it is an observation of the darkening of the disk of the host star during transit. The second method is Doppler spectroscopy, which registers the modulation of the spectral lines because of the center-of-mass motion of the star system. Doppler spectroscopy was historically the first method, and, naturally, the first observations were large planets close to their star, i.e., "Hot Jupiters," hardly conducive to habitability. Thirdly, it is the transit-timing method, which analyses anomalies in the motion of other exoplanets in the system. The fourth method is microlensing – the imaging of the planetary system using gravitational lensing of the star (Fischer D. A. 2014), (Bozza 2016).

All these methods, except microlensing, involve the planet's back-action on the host star's system. However, these methods would be difficult to implement even for the distant planets of the Solar System. On the contrary, phase diagnostics is limited only by instrumental considerations (Shouten 2003), (Cotte 2011).[2] That is easy to understand given the fact that the phase of an electromagnetic field has a logarithmic singularity near each point in space where $|\vec{E}(\vec{r})| = 0$. Heretofore, the phase derivative at this point is infinite, and the sensitivity close to this point is greatly enhanced.

---

[2] Shouten (Shouten 2003) et op. cit. described the phase singularities with infinite derivatives but did not present a measurement procedure leading to superresolution. Currently, the superresolution of phase images is widely used in biological imagery.

On Earth, we obtain only the amplitude distribution of astronomical objects because of the gigantic phase obtained by the starlight during its propagation. However, intensity interferometry with a photon count uses quantum properties of light to eliminate a common phase shift of the photons propagating from a star or reflected from a planet. Diverse applications of the Brown-Twiss interferometry were reviewed by (Baum 1997). His review paper emphasized that the Brown-Twiss detection rate is expressed through a Fourier transform of the density distribution.

We considered a uniformly lit system with the x-axis along the line connecting the centers of a star and a planet, and the observation in the direction of the star's center. The amplitude of the Hanbury-Brown-Twiss (HBT) signal from an entire system is not sufficiently different from that of an isolated star. However, (Ireland 2014) when we register the quadratures of the signal in the selected geometry, the cosine quadrature reproduces the signal from the star alone. In contrast, the sine quadrature emphasizes the difference between the Fourier image of the star and the planet.

2. **An outline of the method**

Diverse applications of the HBT interferometry were reviewed in (Baum 1997). The method uses coincidence in the counts of two remote detectors as a signal, which reflects the intensity correlation of an object. In his review paper, Baum demonstrated that the HBT detection signal could be expressed through a Fourier transform of the density distribution:

$$C(\vec{q}) - 1 = \left| \int \rho(\vec{r}) e^{i\vec{q}\cdot\vec{r}} d^3 r \right|^2 \tag{1}$$

In Equation (1), $\rho(\vec{r})$ is a three-dimensional density distribution of the luminous object, $\vec{q} = \vec{k}_1 - \vec{k}_2$ is a spatial frequency, $\vec{k}_i = k \cdot \vec{r}_i$, are the wavevectors in the direction of the detector $i$=1,2, and $C(\vec{q})$ is an intensity correlation coefficient:

$$C(\vec{q}) = \frac{<I_1 I_2>}{<I_1><I_2>}. \tag{2}$$

In Equation (2), $I_1$ and $I_2$ are the intensities in two channels of the interferometer. Gaussian random field approximation for the averages is almost always justified for astrophysical measurements (Goodman 2007), but can be suspect in biology where HBT begins to be applied (Cotte 2011).

Naturally, in the case of astronomical observation, the density distribution is effectively two-dimensional because we observe a projection of an object on a focal plane of the telescope(s).

Equation (1) already suggests the reasoning behind the method. Namely, if the intensity of the signal from a star is proportional to $|E_{star}|^2$ and the signal from the planet is proportional to $|E_{planet}|^2$, then, the intensity of the HBT signal due to the presence of the planet in the largest order is proportional to their product, $<E_{star} E^*_{planet}>$. Hence, the magnitude of the sine quadrature is proportional to $\sqrt{I_{star} I_{planet}}$.

As an example, we consider a uniformly lit object of the following configuration presented in Fig. 1:

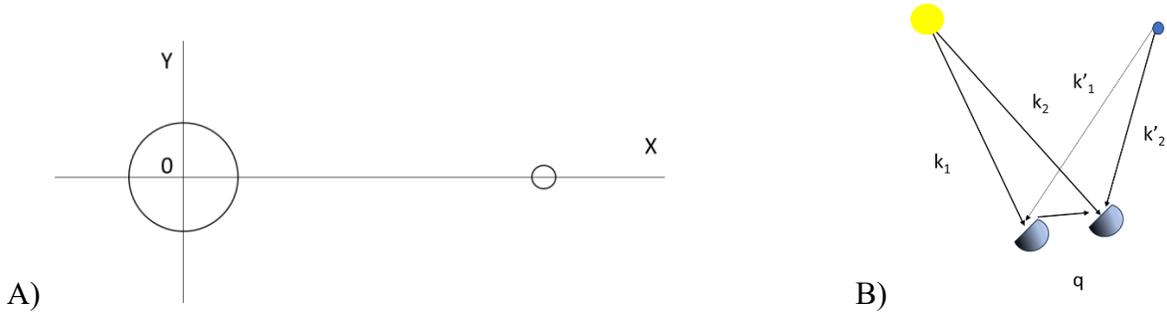

**Fig. 1** A) The geometry of the image (not in scale). Observation is directed at the center of the star. The planet is located along the X-axis, B) The directions of wave vectors. The current resolution of the angular distance between exoplanets and their stars is typically $10^{-3} - 10^{-4}$ *mas* (milliarcseconds).

For demonstration purposes, our plots correspond to a model planetary system with the following characteristics. The planet has a 1/10 radius of the host star, a distance of 10 diameters from the host star, and a luminosity of $10^{-4}$ of a star, which presumes an albedo close to unity and negligible thermal radiation of the planet itself. In practice, the luminosity of the exoplanets is typically eight to ten orders of magnitude smaller than that of their star, but we make our numerical simulations for illustrative purposes only [7]. We provide a possible sketch of the observational setup in Fig. 2.

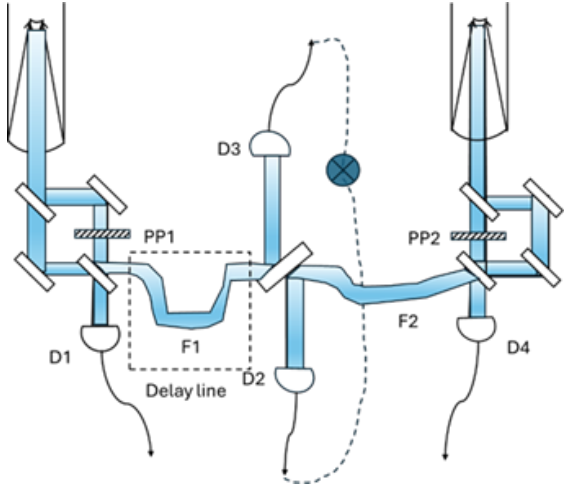

Fig. 2 Binocular telescope installation. D1-D4 are the detectors, PP1,2 are the phase plates, and F1,2 are the fiber optic connections to the interferometer. The coincidence counter between detectors D2 and D3 is shown as a crossed circle. An additional mode-locking cavities may be required. Also, the electronic coincidence counter may be replaced with the fiber optic correlator (Guyon 2002).

Once the photon counts in the interferometer's two arms are balanced and their stationarity established, one can independently measure the intensity fluctuations in each arm and the coincidence count between the arms. According to the results of Section 3 and Equations (4.3-5 and 4.3-(23-25)) of Mandel and Wolf, these measurements can establish the quadratures (Mandel 1995). Orthogonal quadratures can confirm or deny the existence of an exoplanet, practically, whether the luminous object is circularly symmetric or not. The advantage of the method is that it seems extremely sensitive to the asymmetry of the target despite a great disparity in luminosities.

3. **Quadratures as signatures of a star-planet system**

The intensity of the signal from the system star+planet (S+P) is not sufficiently different from the star's signal alone (see Fig. 4 B). However, because of the linearity of the Fourier transform, one can express the factor in Equation (1) as:

$$\left|\mathcal{F}_{\vec{r}\to\vec{q}}(\rho(\vec{r}))\right|^2 = C^2_{S+P}(\vec{q}) + S^2_{S+P}(\vec{q}) \qquad (3)$$

With slight abuse of terminology and notation, we call the real and imaginary parts of the Fourier transform the quadratures, notably the cosine and sine quadrature:

$$\begin{aligned} C_{S+P} &= Re\big(\mathcal{F}_{star}(\vec{q}) + \mathcal{F}_{planet}(\vec{q})\big) \\ S_{S+P} &= Im\big(\mathcal{F}_{star}(\vec{q}) + \mathcal{F}_{planet}(\vec{q})\big) \end{aligned} \qquad (4)$$

or, in an expanded form:

$$C_{S+P} = |\mathcal{F}_{star}(\vec{q}) + \mathcal{F}_{planet}(\vec{q})| \cdot Cos[Arg(\mathcal{F}_{star}(\vec{q}) + \mathcal{F}_{planet}(\vec{q}))]$$
$$S_{S+P} = |\mathcal{F}_{star}(\vec{q}) + \mathcal{F}_{planet}(\vec{q})| \cdot Sin[Arg(\mathcal{F}_{star}(\vec{q}) + \mathcal{F}_{planet}(\vec{q}))]' \quad (4')$$

where $\mathcal{F}(.)$ is the Fourier transform of the luminous density profile.

In the chosen geometry of Fig. 1, the cosine quadrature reproduces the signal from the star alone (Fig. 2). The sine quadrature emphasizes the difference between the Fourier image of the star and the planet. The image in Fig. 3 confirms this intuition. Unless the orbit lies in the (X, Z) plane, where direction Z is perpendicular to the focal plane, observed signals will be mixtures of the quadratures of Equation (3).

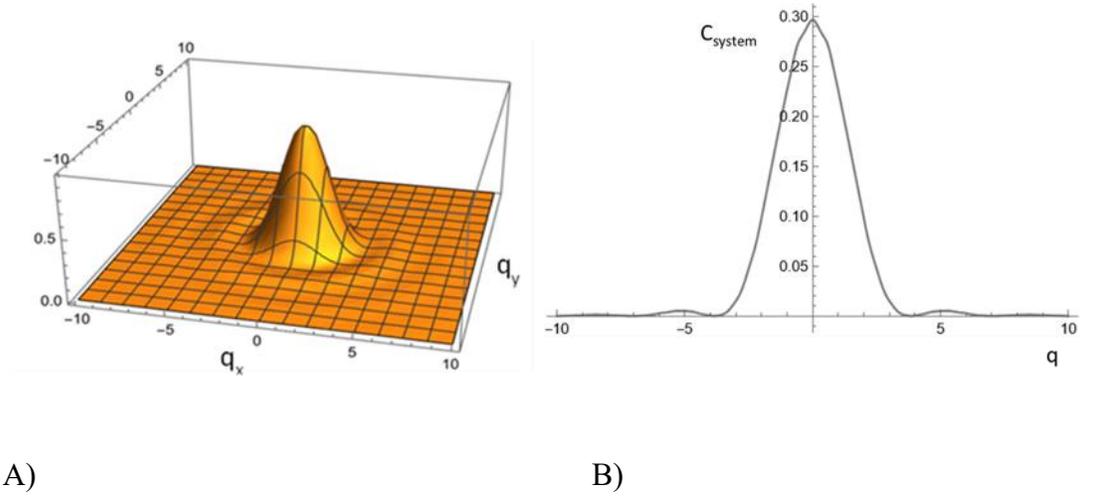

A)                                     B)

**Fig. 3** A) Two-dimensional projection of the cosine quadrature of the Brown-Twiss signal (Equation 4). $q_x$ and $q_y$ are the spatial frequencies in the observation plane. B) Cross-section of the intensity plot. The difference with the signal from the star itself is barely noticeable.

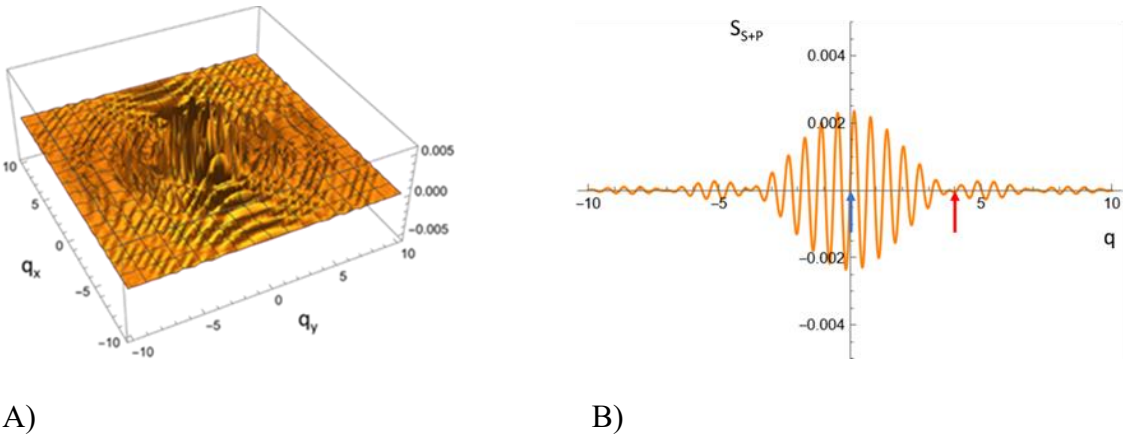

A)                                     B)

**Fig. 4** Sine quadrature of the HBT signal from the planetary system of Fig. 1. If a lone star is projected in the middle of the focal plane, the sine quadrature is identical to zero. A) Two-dimensional projection. B) Cross-section plot as a function of $q = \sqrt{q_x^2 + q_y^2}$ in the direction of $q_x$ ($q_y$=0). Sharp oscillations indicate a possibility of achieving high experimental signal-to-noise. Blue and red arrows designate a near-optimal measurement setting for two telescopes when the second telescope is adjusted to the null signal.

In that case, a Fourier transform of a planet's image can be expressed as a shifted Fourier transform of the centered planet's image. Fig. 4 shows that the sine quadrature displays sharp peaks with discontinuous derivatives caused by the argument of the complex Fourier amplitude changing sign.

$$\tilde{C}_{S+P} = \left|\mathcal{F}_{star}(\vec{q}) + e^{iq_x a}\tilde{\mathcal{F}}_{planet}(\vec{q})\right| \cdot Cos\left[Arg\left(\mathcal{F}_{star}(\vec{q}) + e^{iq_x a}\tilde{\mathcal{F}}_{planet}(\vec{q})\right)\right]$$
$$\tilde{S}_{S+P} = \left|\mathcal{F}_{star}(\vec{q}) + e^{iq_x a}\tilde{\mathcal{F}}_{planet}(\vec{q})\right| \cdot Sin\left[Arg\left(\mathcal{F}_{star}(\vec{q}) + e^{iq_x a}\tilde{\mathcal{F}}_{planet}(\vec{q})\right)\right]$$
(5)

In Equation (4), the variable with a tilde is a Fourier transform of a planet's distribution shifted to the origin, and *a* is the difference in optical paths in channels 1 and 2. As in Fig. 1, we presume that the telescope is directed to the center of the star.

The quadrature method is based on two simple facts. First, for a 2-D Fourier transform and circularly symmetric distribution of an object, one can always find a coordinate system in which it is real. This cannot be done, in general, if the circular symmetry is absent. The second fact is that the modulus and the cosine of the correlation coefficient can be observed simultaneously.

4. **Determination of the correlation by the Max-Zehnder interferometer**

Amplitude interferometry of the star image can determine the correlation coefficient of the incoming electric field of the star system. The channels of the interferometer are shown in Fig. 2. A stylized setup is shown in Fig. 5.

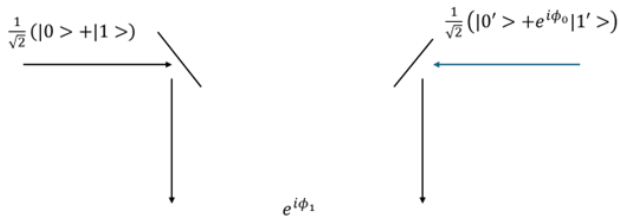

A)

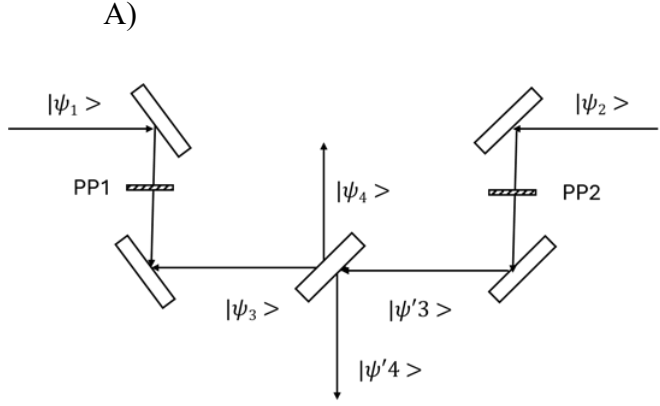

B)

Fig. 5 A) Interferometric setup for measuring the cosine of the coherence function. Channels of the interferometer. B) Interferometric setup for measuring the cosine of the coherence function. Transformation of the wave function of the photon inside the interferometer.

The wave function of the incoming photon in the first and second channels is as follows:

$$|channel_1> = \frac{1}{\sqrt{2}}(|0> + |1>) \qquad (6)$$

$$|channel_2> = \frac{1}{\sqrt{2}}(|0'> + e^{i\phi_0}|1'>) \qquad (6')$$

where $\phi_0 = k \cdot d$ is the phase difference between the interferometer arms. We presume that the channels are otherwise symmetric. The symbols $|0>$ and $|1>$ are the wave functions of photons with 0 ('vacuum') and 1 photon, respectively. Because of the extremely feeble images of the stars, the presence of more than 1 photon in the interferometer per coherence time of the starlight is highly improbable, and we do not need to consider multi-photon states of radiation.[3]

For the $|\psi_{1,2}>$, the following equations must hold:

$$|\psi_1> = \frac{1}{\sqrt{2}}\left(\frac{1}{\sqrt{2}}(|0> + |1>) + \frac{1}{\sqrt{2}}(|0'> + e^{i\phi_0}|1'>)\right) \qquad (7)$$
$$|\psi_2> = \frac{1}{2}\left((|0> + |1>)e^{-i\phi_1} + (|0'> + e^{i\phi_0}|1'>)\right)$$

In the first of Equations (7), we explicitly retained normalization factors. A beam splitter can be described by the matrix:

$$\begin{pmatrix} \alpha & \beta \\ -\beta^* & \alpha^* \end{pmatrix} \qquad (8)$$

---

[3] Assuming energy flux of $10^{-27}$ W·Hz$^{-1}$/m² (a weak star case in (de Graauw 1982)) for 2×78 m² collection area of the telescopes, $\Delta v_s = 6 \cdot 10^{13}$ Hz, and $\tau_{coh} = 1.5 \cdot 10^{-13}$ sec. for blackbody photons, we arrive at the estimate of $1.1 \cdot 10^{-6}$ photons during coherence time.

where for a non-absorbing splitter, $|\alpha|^2 + |\beta|^2 = 1$. Heretofore, we obtain the set of equations for the wavefunctions $|\psi_{3,4}>$ and $|\psi'_{3,4}>$:

$$|\psi_3> = \frac{\alpha}{2}\left((|0> +|1>) + (|0'> +e^{i\phi_0}|1'>)e^{i\phi_1}\right)$$
$$|\psi'_3> = \frac{\beta}{2}\left((|0> +|1>)e^{i\phi_1} + (|0'> +e^{i\phi_0}|1'>)e^{i\phi_1}\right)$$
(9)

And

$$|\psi_4> = -\frac{\beta^*}{2}\left((|0> +|1>) + (|0'> +e^{i\phi_0}|1'>)e^{i\phi_1}\right)$$
$$|\psi'_4> = \frac{\alpha}{2}\left((|0> +|1>)e^{i\phi_1} + (|0'> +e^{i\phi_0}|1'>)e^{i\phi_1}\right)$$
(10)

Because vacuum states in each channel are orthogonal – they essentially represent a shot noise, the emergence of mutually coherent photons simultaneously in both channels is highly improbable.[4] We can write down the expressions for the expectation probabilities in the channels:

$$\begin{aligned}<0|0'> &= 0 \\ <1|1'> &\approx 0 \\ <0|1'> &= \cos(\gamma_{12}) \\ <1|0'> &= \cos(\gamma^*_{12})\end{aligned}$$
(11)

The last two formulas follow from a definition of the coherence function (Mandel 1995). So, we obtain a formula for the signal in the lower detector of the interferometer in Figs. 2 and 5.

$$<\psi_3|\psi_4> = -\alpha\beta^*[\cos(\phi_0) + \cos(\phi_1) + 2\cos(\phi_0 - \phi_1)\cos(Re[\gamma_{12}])] \quad (12)$$

A similar expression can be derived for the upper detector [check primed and non-primed channel notation with respect to each detector – P. L. !]. By scanning through the phase delays in the arms of the interferometer $\phi_0$ and $\phi_1$, one can determine the cosine of the correlation coefficient as a function of the visibility of interference fringes.

---

[4] A typical width of the observed spectrum is 1/5 to 1/100 of the maximum wavelength (Gardner J. P. 2008). A crude estimate of the coherence time of the 2-micron radiation by the formula $\tau_c \cong \frac{\lambda^2}{\Delta\lambda c}$ gives the value between three and six times $10^{-13}$ sec. For the flux typical for a weak star (de Graauw 1982), the average number of photons detected for a unit quantum efficiency during the coherence time is several times $10^{-7}$.

5. **Comparison of the signal-to-noise ratios of Michelson interferometry, Hanbury-Brown-Twiss interferometry, and Heterodyne interferometry**

It is well known that the SNR of the Hanbury-Brown-Twiss interferometry is low compared to the amplitude interferometry, also commonly called Michelson interferometry, independently of the setup of the interferometer itself (Michelson, Sagnac, or echelle). However, the quantitative comparison of the SNR for the different interferometry types is provided in low-circulation or old editions, in different units, and we find it useful to summarize it here.

**Table 1** Signal-to-noise ratios for the select interferometric setups (from T. de Graauw and H. van de Stadt, Table II (de Graauw 1982)). All the interferometric schemes in the case of the weak signal measure the modulus square of the coherence function.

| Type | SNR |
|---|---|
| HBT | $\dfrac{F_s^2 \Delta \nu}{\sqrt{2}\,(NEP)^2_{inc}} \sqrt{\Delta f \cdot T_0}\; |\gamma_{12}(d)|^2$ |
| Michelson, strong signal | $\dfrac{F_s \Delta \nu_s}{(NEP)_{inc}} \sqrt{T_0}\, |\gamma_{12}(d)|$ |
| Michelson, weak signal | $\dfrac{F_s^2 \Delta \nu_s^2}{(NEP)^2_{inc}} \sqrt{\dfrac{T_0}{B_d}}\; |\gamma_{12}(d)|^2$ |
| Heterodyne, strong signal | $\dfrac{F_s \sqrt{B_{IF}}}{(NEP)_{coh}} \sqrt{T_0}\, |\gamma_{12}(d)|$ |
| Heterodyne, weak signal | $\dfrac{F_s^2 B_{IF}}{(NEP)^2_{coh}} \sqrt{\dfrac{T_0}{B_d}}\; |\gamma_{12}(d)|^2$ |

In Table 1, $T_0$ is the observation time, $\Delta f$ is the electrical bandwidth, $B_d$ is the detector bandwidth, $\Delta \nu$ is the width of a stellar spectrum, $\Delta \nu_s$ is the inverse coherence time for this spectrum, and $B_{IF}$ is the width of a sideband. $F_s$ is the spectral power of a signal $[W \cdot Hz^{-1}]$. $NEP_{coh}\ [W \cdot Hz^{-\frac{1}{2}}]$ is a Noise Equivalent Power for the detector, and $NEP_{inc} = k_B T_Q\ [W \cdot Hz^{-1}]$ is a Noise Equivalent Temperature $T_Q$ for the detector measured in units of the Boltzmann constant.[5] Finally, $\gamma_{12}(d)$ is the required observable – the correlation coefficient measured for the distance $d$ between the detectors.

By simple rearrangement of the expressions in Table 1, we arrive at the comparative expressions for the ratios of the SNR factors.

**Table 2**. Ratios of the SNR of the different types of interferometry.

| Type | Ratio |
|---|---|
| HBT: Michelson, weak signal | $\dfrac{\Delta\nu}{\Delta\nu^2{}_s}\sqrt{\Delta fB_d} \lesssim \dfrac{\sqrt{\Delta fB_d}}{\Delta\nu_s} \ll 1$ |
| Michelson, weak signal : Heterodyne | $\dfrac{\Delta\nu_s{}^2}{B_{IF}} \cdot \dfrac{(NEP)^2{}_{coh}}{(NEP)^2{}_{inc}} \gg 1$ |
| Heterodyne, weak signal: HBT | $\dfrac{B_{IF}}{\Delta\nu\sqrt{\Delta fB_d}} \dfrac{(NEP)^2{}_{inc}}{(NEP)^2{}_{coh}} - ?$ |

6. **Estimation of the scale of the interferometer**

For determinacy, we assume that the observed system is a planet orbiting at 5 a.u. a red dwarf with 0.2 of the solar radius and T=2500 K, which is located at 100 parsecs. Then, the coherence radius for the entire system – we assume that the star and planet images are unresolved– will be around 4.5 m, and the coherence radius for the star alone will be 5 km. Thus, a realistic scale of "fast" and "slow" oscillations outlined in Figure 4B is 1000:1. It demonstrates the difficulty of using Michelson interferometry, which is extremely sensitive to the absolute accuracy of the path difference (fractions of the wavelength) for the measurement of slow oscillations of the quadrature. This is not impossible given the success of the LIGO and VIRGO interferometers (Collaboration 2015). However, it can increase the cost of the installation by several orders of magnitude and cancel the advantages of the proposal.

Note that the zero point of a slow oscillation of the quadrature corresponds to the distance between the detectors, where we eliminated the influence of the star altogether! Yet, the amplitude of the fast oscillation near this point is exceedingly small. During an actual measurement, it is advisable to keep one telescope at a null point and change the relative distance of the other detector until the maximum magnitude of the quadrature is achieved.

---
[5] By this convention, we correct a frequent notation error first noticed by the author of (Scheider 2014) (footnote [9]).

7. **Conclusion**

Our paper proposes a method of exoplanet discovery inspired by the Hanbury-Brown-Twiss interferometry. This method is particularly suitable for discovering exoplanets, which are distant from the host star and influence it only slightly. This method relies only on the asymmetry of the focal image.

HBT and heterodyne interferometry measure $|\gamma(\vec{r},\vec{r}')|^2$, the square modulus of the correlation function. Amplitude interferometers at each of the observation sites measure $Re[\gamma(\vec{r},\vec{r}')]$. Knowing $|\gamma(\vec{r},\vec{r}')|$ and $Re[\gamma(\vec{r},\vec{r}')]$, we can effectively determine the phase $\phi(\vec{r},\vec{r}')$ of the correlation function: $\gamma(\vec{r},\vec{r}') = |\gamma(\vec{r},\vec{r}')|e^{i\,\phi(\vec{r},\vec{r}')}$.[6] Resolution of the phase of the correlation function does not need to obey a Rayleigh limit and is exceptionally sensitive to the asymmetry of a luminous system.

An increase in resolution concerning the Rayleigh criterion is possible because the fluctuations in the photo-current depend on the correlation between distant detectors. While the correlation between distant detectors is perfectly classical (Mandel 1995), Ch. 4; the gain in resolution can be obtained only through the photon counting methods.

The measurement of the quadrature can be accomplished through a binocular setting of two telescopes. A required dynamic range of the telescopes in the pair is assumed to be between the correlation radius $r_c \sim \lambda/\Delta\theta_{s+p}$ of the entire planetary system and the correlation radius $r_c \sim \lambda/\Delta\theta_s$ of the luminous star. Here $\lambda$ is the observation wavelength, $\Delta\theta_s$ is the angular dimension of the star, and $\Delta\theta_{s+p}$ – is the angular dimension of the prospective system. For an imaginary system of the five a.u. size at the distance of 10 parsecs illuminated by a dwarf star with 20% of the diameter of the Sun, observed at 1.2 µ, a requisite dynamic range will be between 4.5 meters and 5 kilometers. Alternatively, using a 100 GHz submillimeter array, the dynamical range increases to 1000 m and 6,600 km. These distances between detectors are accessible through Very Large Baseline Interferometry (VLBI, (Bowman 2016), (Fish 2019)). From the above estimate, it seems that a near-to-mid infrared range of observation within the atmospheric window with a dynamic range between meters and kilometers might be optimal.

---

[6] In a proposed experiment, the observables are the cosine and sine quadrature, which combine both the amplitude and phase of an incoming light (see Equations (4) and (5)).

**Acknowledgements**. The author expresses deep gratitude to Alexandre Mayer (U. Namur), who participated in the first version of the paper (Lerner 2023), for his advice, and to James Fienup (U. Rochester) for the suggestions for the next development.

**Note on code and data availability**: The author's notebooks in *.nb are available from the corresponding author on reasonable request.

Appendix. **Phase superresolution**

The formula for the spatial resolution of amplitude optical devices was developed by Rayleigh and Airy (Born 1999). Quantitative expression of the Rayleigh criterion in angular units is as follows:

$$\theta_m/2 = 0.61 \frac{\lambda}{D} \qquad (A.1)$$

Where $\lambda$ is the wavelength and D is the aperture.

The resolution of the phase-registering optical device does not obey the Rayleigh criteria. Close to the null of zero amplitude $|\vec{E}(\vec{r})| = 0$, the phase has a logarithmic singularity (Goodman 2007). Logarithmic singularity of the optical phase results in 1/r singularity of the responsivity to the small deviations near the null. Phase resolution of the distance can be arbitrary and is limited only by the signal-to-noise ratio of the optical device. Below, we provide a simple argument in favor of this statement. More detailed reasoning can be found in (Tychinskii 2008).

There is no simple generalization of uncertainty principle for the photon number-phase relationship because the phase operator for the optical field does not exist in an ordinary sense. However, operators of trigonometric functions for the phase can be consistently defined (Loudon 2000) [in reality, Louisell, and Carruthers and Nieto – P. L.]. Using Loudon's definition, we have the following uncertainty relations:

$$<\Delta N><\Delta sin^2\varphi> \geq \frac{1}{2}, \quad <\Delta N><\Delta cos^2\varphi> \geq \frac{1}{2}. \qquad (A.2)$$

In Equation (A.1), the expressions in brackets are expectations of variances of the quantum operators for the number and the squares of the sine and cosine of phase, respectively. For a practical interferometer, the phase is reasonably well defined, and its uncertainty is small. So, we can assume that the classical phase resolution $<\Delta\varphi^2>_{pr}$ is equal to:

$$<\varphi|\Delta sin^2\varphi|\varphi> \approx <\Delta\varphi^2>_{pr} \qquad (A.3)$$

A minimum energy in an optical mode is $\hbar\omega$ per mode (we consider the field monochromatic for simplicity). If the total number of modes in the device is $m_r$, the uncertainty relations (A.1), and Equation (A.2) can be rewritten in the form:

$$\hbar\omega <\Delta N><\Delta\varphi^2>_{pr} \geq \hbar\omega m_r = N_{min}, \qquad (A.4)$$

Where $N_{min}$ is the minimum noise. For the resolvable signal, $<\Delta N> \leq <N>$ where $N$ is a photon number. If we approximate the signal as $S = \hbar\omega <N>$, and assume the equality sign in the minimum, Equation (A.3) can be rewritten as

$$S \geq \hbar\omega <\Delta N><\Delta\varphi^2>_r = N_{min} \qquad (A.5)$$

For the resolvable angle, we receive the inequality:

$$\sqrt{<\Delta\varphi^2>_r} \geq \left(\frac{N_{min}}{S}\right)^{1/2} \qquad (A.6)$$

Heretofore, the resolution of the phase-detecting devices can be larger in proportion to the square root of the:

$$\frac{\theta_m}{\sqrt{<\Delta\varphi^2>_r}} = 2\left(\frac{S}{N_{min}}\right)^{1/2} \qquad (A.7)$$

More accurate estimates are provided in (Tychinskii 2008).

To estimate the signal-to-noise ratio, we use the ratio between absolute values of the logarithmic derivatives of Fourier spectra of the signal and of the noise, respectively ((Tychinskii 2008), Fig. 1c). The intuition behind this estimate is that a fundamental noise is produced by the photon coming directly from a star and the signal – from a photon emitted by a star and reflected by a planet. Heretofore, the maximum signal-to-noise ratio can be estimated as a ratio of responses to the photon coming from a planet to the photon coming from a star. A signal-to-noise ratio is plotted for the imaginary geometry of Fig. 1 and Section 2 in Figure 6. Because the amplification factor is proportional to the square of the S/N ratio, the expected maximum enhancement of the resolution is on the order of ~70.

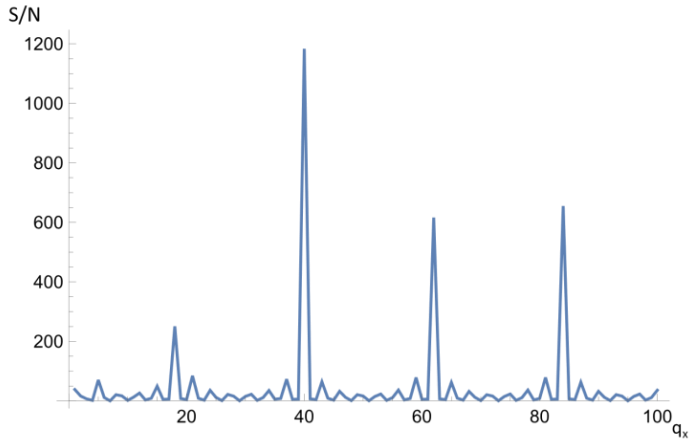

Fig. 6. The estimated signal-to-noise ratio of the "Star+Planet" system from Fig. 1 as a function of spatial frequency.